\documentstyle[12pt,openbib]{article}
\title{Rindler and Minkowski particles relationship revisited}
\author{George E.A. Matsas\\
         Instituto de F\'\i sica Te\'orica\\ 
         Universidade Estadual Paulista\\
         Rua Pamplona 145\\
         01405-900-S\~ao Paulo, S\~ao Paulo\\
         Brazil}

\begin{document}
\maketitle
\begin{abstract} 
We show that the emission of a Minkowski particle by a general 
class of scalar sources as described by inertial observers corresponds 
to either the emission or the absorption of a 
Rindler particle as described by uniformly accelerated observers.
Our results are discussed  in connection with the current controversy 
whether uniformly accelerated detectors radiate. 

04.62.+v, 04.70.Dy
\end{abstract}
\newpage

There is currently a discussion about what is the influence 
that accelerated detectors have on the quantum field to which 
it is coupled \cite{Gr}--\cite{AM}. Raine et al \cite{RSG} 
and Massar et al \cite{MPB} have argued  that a uniformly accelerated 
oscillator does not radiate. This is in contrast to 
Unruh and Wald's result \cite{UW} that the absorption of a 
Rindler particle by a uniformly accelerated
Unruh-DeWitt detector \cite{U}-\cite{D} corresponds to the emission
of a Minkowski particle. This controversy recalls the classical 
discussion whether uniformly accelerated electric charges radiate 
\cite{FR}. The challenge consisted in understanding how 
the standard formula which predicts that a charge radiates
whenever its acceleration does not vanish could be compatible with
the fact that the radiation reaction associated with 
uniformly accelerated trajectories vanishes.
It is known, however, that no paradox appears provided one
considers electric charges following  {\em realistic trajectories}.
By realistic trajectories we mean trajectories not only everywhere
timelike, but also asymptotically timelike. 
In other words, it is experimentally and theoretically 
free of dispute that 
charges following non-inertial asymptotically-timelike  
trajectories radiate. This motivates us to consider the
radiation emitted by detectors following more realistic trajectories 
than uniformly accelerated ones.  Here we show that  the emission  
(by a general class of scalar sources following {\em arbitrary} 
worldlines) of 
a Minkowski particle in the inertial vacuum as 
described by inertial observers  corresponds 
to the thermal emission or absorption of a Rindler particle 
as described by uniformly accelerated observers.
We argue that this result must be reliable 
for any sources following realistic trajectories. 
One such a class of sources are the so called finite--time 
detectors, i.e. detectors kept switched on for a finite 
amount of time \cite{HMP}.   
We consider our approach conceptually simpler 
than the one developed in Refs. \cite{UW}--\cite{U} 
in the sense that it does not rely explicitly
on the properties of the field on the Killing horizon. 
We hope this paper sheds some light on the controversy above.
We assume natural units  $(\hbar=c=k_B=1)$, and an n-dimensional 
Minkowski spacetime  with signature $(+ - - ... -)$. 

As explained above, we will be interested here in 
analyzing the radiation emitted by detectors following
arbitrary trajectories rather than only hyperbolic ones.
In a first step we will discuss point--like
detectors following arbitrary worldlines 
(see also Ref. \cite{AMH}), and
in a second step we will broader our discussion in order to 
consider eventually the radiation emitted 
by more general quantum devices rather than only point--like ones. 
A point--like detector is
represented by a monopole $\hat m(\tau )$. We couple it to a real massless
scalar field  $\hat \phi [x^\mu (\tau )]$ where $x^\mu (\tau )$
is the detector's worldline, and $\tau$ is its proper time. The interaction
action between the detector and the field will be
\begin{equation}
\hat {\cal S}_I = \int_{-\infty}^{+\infty} 
                  d\tau c(\tau ) \hat m (\tau )
                  \hat \phi [x^\mu (\tau )] ,
\label{1}
\end{equation}
where $c(\tau )$ is a small coupling constant which can be used to switch on
and off the detector. In the Heisenberg picture the 
monopole operator is time--evolved 
by $\hat m(\tau) =e^{i H_0 \tau} \hat m(0) e^{-i H_0 \tau}$,
where $ H_0$ is the free Hamiltonian of the detector. In the first order
of perturbation theory, the transition amplitude for the detector to
jump from the energy eigenstate $\vert E \rangle$ to $\vert E' \rangle$,
and simultaneously emit a particle is
$
{\cal A}_{\mbox{\scriptsize em}} = 
              \langle  1 \vert \otimes \langle E' \vert 
              \hat{\cal S}_I  
              \vert E \rangle \otimes  \vert 0 \rangle .
$
Since we are not interested here on the back-reaction that the emission 
of a particle produces on the detector itself, it is interesting for our
purposes to rewrite this amplitude as 
\begin{equation}
{\cal A}_{\mbox{\scriptsize em}} = 
              \langle  1 \vert \hat{\cal S}  \vert 0 \rangle ,
\label{2}
\end{equation}
where
\begin{equation}
\hat {\cal S}  = \langle E' \vert \hat {\cal S}_I \vert E \rangle .
\label{3}
\end{equation} 
By using (\ref{1}), we can calculate (\ref{3}) explicitly
\begin{equation}
\hat {\cal S} = \int_{-\infty}^{+\infty} 
                  d\tau c(\tau ) 
                 \langle E' \vert \hat m ( 0 ) \vert E \rangle
                  e^{i \Delta E \tau} \hat \phi [x^\mu (\tau )] ,
\label{4}
\end{equation}
where $\Delta E=E'-E$ is the energy gap of the detector.
For sake of our further generalization, Eq. (4) will be rewritten 
in a more useful form as
\begin{equation}
\hat{\cal S} = 
\int d^n x \sqrt{\vert g(x) \vert } \; j(x) \hat \phi (x) ,
\label{S1}
\end{equation}
where for this point--like detector following the worldline 
$x^\mu = x^\mu (\tau )$, we have
\begin{equation}
j(x') = c (\tau ) \langle E' \vert \hat m(0) \vert E \rangle e^{i\Delta E \tau}
     \delta^{n-1} [{\bf x}' - {\bf x} (\tau)] / \sqrt{\vert g(x') \vert } u^0 .
\label{6}
\end{equation}
Notice that $j(x)$ is clearly complex. The 
$e^{i \Delta E \tau}$ factor in (\ref{6})  simply reflects the detector's
transition, while the delta function
appears because at this point we are restricted to point--like devices.
In order to extend our analysis to more general quantum devices 
than point--like ones, we will
assume throughout the rest of the paper that $j(x)$ in (5) is a general 
complex c-number, rather than having some particular form like (\ref{6}). 
(Clearly, one can eventually particularize 
our results to point--like detectors by 
substituting (\ref{6}) in our final formulae as briefly 
commented at the end, or particularize our results to classical sources
where $j(x)$ is real as comprehensively studied in Ref. \cite{HM}.)   
  
Concluded the preliminaries, 
let us begin by computing the emission rate of Minkowski particles
as described by inertial observers. 
Our strategy will consist in relating it with the emission and absorption
rates of Rindler particles as described by uniformly accelerated observers. 
For this purpose we will deal with the explicit solution of the field
in both frames, rather than relying on the properties of the field on
the Killing horizon. In the first order of perturbation theory, 
the emission amplitude of a Minkowski particle as
calculated in the inertial frame is [see Eq. (\ref{2})]
\begin{equation}
^M{\cal A}_{\mbox{\scriptsize em}}
(k,{\bf k}) = 
              _M\langle  {\bf k} \vert \hat{\cal S}  \vert 0
\rangle_M ,
\label{MAEM}
\end{equation}
where $k= \vert {\bf k} \vert$, and $\hat {\cal S}$ is given by 
Eq. (\ref{S1}).
Expanding the scalar field $\hat \phi(x)$ 
in terms of positive and negative energy modes with respect to
inertial observers \cite{BD} 
\begin{equation}
\hat \phi (x^\mu ) = \int \frac{d^{n-1} {\bf k}}{ \sqrt{2 k (2 \pi)^{n-1}}}
(\hat a_{\bf k}^M e^{-ik_\mu x^\mu} + {\rm H.c.} ),
\label{FIXMU}
\end{equation}
we express from (\ref{S1}) the emission amplitude (\ref{MAEM}) as  
\begin{equation}
^M{\cal A}_{\mbox{\scriptsize em}}
(k,{\bf k}) = \left[ 2 k (2 \pi)^{n-1}\right]^{-1/2} \int d^n x
j(x) e^{i k_\mu x^\mu} ,
\label{MAEM1}
\end{equation}
where $x^\mu \equiv (t,x,y^{2\leq i \leq n-1})$.
Thus, the emission probability of a Minkowski particle as described by
inertial observers is 
\begin{equation}
^M{\cal P}_{\mbox{\scriptsize em}}
= \int d^{n-1} {\bf k} \,
\vert ^M{\cal A} {\mbox{\scriptsize em}(k, {\bf k} )} \vert^2 ,
\label{PM}
\end{equation} 
where $^M{\cal A}_{em}$ is given in (\ref{MAEM1}). 

Next, we aim to express the  emission rate of Minkowski
particles (\ref{PM}) in terms of the  emission and  absorption
 of Rindler  particles. For this purpose, it is convenient to introduce
Rindler coordinates $(\tau, \xi, y^{2 \leq i\leq n-1})$. 
These coordinates are related with Minkowski coordinates
$(t,x,y^{2\leq i\leq n-1})$  by
\begin{equation}
t = \frac{e^{a\xi}}{a}\sinh a\tau, \;
x = \frac{e^{a\xi}}{a}\cosh a\tau.
\label{RC}
\end{equation}
A worldline defined by $\xi, y^{2\leq i\leq n-1} =$ const describes an
observer with constant proper acceleration $a
e^{-a\xi}$. We will denominate these observers {\em Rindler} observers.
The natural manifold to describe Rindler observers
is the Rindler wedge, {\em i.e.} the portion of the Minkowski space defined
by $x > \vert t \vert$. It is crucial for our purposes that the current
$j(x^\mu)$ be confined inside this wedge, otherwise
it will not be able to be analyzed by accelerated observers.
This is the reason why the question {\em whether inertial charges
emit or not radiation with respect to uniformly accelerated observers}
is not well posed in general. (Notwithstanding, see \cite{M} for a 
full solution in a particular case.) 
The Rindler wedge is a globally hyperbolic spacetime 
in its own right, with a Killing horizon at $x=\pm t$ ($\tau =
\pm \infty $) associated with the boost Killing field $\partial_
\tau$ at its boundary.
Using (\ref{RC}),
the Minkowski line element restricted to the Rindler wedge is
\begin{equation}
ds^2 = e^{2a\xi} (d\tau^2 - d\xi^2) - \sum_{i=2}^{n-1} (d {y^i})^2 .
\label{DS}
\end{equation}
Solving the massless Klein-Gordon equation $\Box \phi =0$ in the
Rindler wedge, we obtain a complete set of Klein-Gordon
orthonormalized functions \cite{F}
\begin{equation}
u_{\omega {\bf k}_\bot} (x^\mu) =\left[ \frac{2 \sinh \pi \omega/a}{
(2\pi)^{n-1} \pi a}\right]^{ \frac{1}{2}} K_{i\omega/a}
\left(\frac{k_\bot}{a} e^{a\xi } \right)
e^{i {\bf k}_\bot {\bf y} - i \omega \tau} ,
\label{U}
\end{equation}
where $k_\bot \equiv \vert {\bf k}_\bot \vert =
\sqrt{\sum_{i=2}^{n-1} (k_{y^i})^2}$, $\omega$  is the frequency
of the Rindler mode, and $K_{i\lambda }(x)$ is the MacDonald function.
Using (\ref{U}), we express the scalar field in the uniformly
accelerated frame in terms of
positive and negative frequency modes with respect to the boost
Killing field $\partial_\tau$  
\begin{equation}
\hat \phi (x^\mu ) = \int d^{n-2} {\bf k}_\bot
\int_{0}^{+\infty} d\omega
\left\{{\hat a^R}_{ \omega {\bf k}_\bot}
       u_{\omega {\bf k}_\bot }
       (x^\mu ) + {\rm H.c.} \right\},
\label{FIXMU2}
\end{equation}
where $\hat a^R_{\omega {\bf k}_\bot}$ is the annihilation operator of
Rindler particles, and obeys the usual commutation relation
\begin{equation}
\left[\hat a^R_{ \omega {\bf k}_\bot } ,
\hat a^{R \dagger}_{ \omega' {\bf k}'_\bot } \right ] =  \delta (\omega -
\omega')
\delta  ({\bf k}_\bot - {\bf k}'_\bot) .
\label{CR}
\end{equation}

It is possible now to Fourier analyze the current $j(x^\mu )$ in
terms of Rindler modes. We define its {\em Rindler--Fourier} 
transform as 
\begin{equation}
\tilde \jmath_R ( \omega_0, \omega, k_\bot) \equiv
\int d^n x \sqrt{\vert g(x)\vert } j(x^\mu ) u^*_{\omega {\bf k}_\bot} 
e^{-i(\omega - \omega_0) \tau} .
\label{RFT}
\end{equation} 
This relation can be easily inverted 
\begin{equation}
j(x^\mu) = \frac{e^{-2a\xi}}{\pi}\int d^{n-2} {\bf k}_\bot
        \int_0^{+\infty} d\omega \omega
        \int_{-\infty}^{+\infty} d\omega_0
\tilde{\jmath}_{R}(\omega_0, \omega , k_\bot ) 
u_{\omega {\bf k}_\bot }(x^\mu ) e^{i(\omega-\omega_0)\tau} 
\label{JX2}
\end{equation}
by using the completeness relation \cite{HM}
\begin{equation}
 \int_0^{+\infty} d\omega \omega \sinh \frac{\pi \omega}{a} \; 
 K_{i\omega/a}\left(\frac{k_{\bot}}{a}e^{a\xi}\right)
K_{i\omega/a}\left(\frac{k_{\bot}}{a}e^{a\xi'}\right) 
=\frac{\pi^2 a}{2}\delta(\xi-\xi') .
\label{CR2}
\end{equation}
In order to relate the particle emission and absorption rates
in both reference frames,  
we substitute (\ref{JX2}) and (\ref{U}) in (\ref{MAEM1}). The
$ y^i$ and ${\bf k}_\bot$ integrals are easy to perform, while 
the $\tau$ and $\xi$ integrals can be solved by  noting that
(use the change of variables $\eta =
e^{a\tau}$ in conjunction with Eq. 3.471.10  of Ref. \cite{GR}) 
\begin{equation}
\int_{-\infty}^{+\infty} d\tau e^{i(kt-k_x x - \omega_0 \tau)} = 
\frac{2}{a} \left[\frac{k+k_x}{k-k_x}\right]^{-i\omega_0/2a}
e^{\pi \omega_0/2a} K_{i\omega_0/a} 
\left( \frac{k_\bot}{a} e^{a\xi} \right) ,
\label{I1}
\end{equation}
and by using the following orthonormality relation \cite{G}:
\begin{equation}
 \int_{-\infty}^{+\infty} d\xi\; 
K_{i\omega/a}\left(\frac{k_{\bot}}{a}e^{a\xi}
\right)
K_{i\omega'/a}\left(\frac{k_{\bot}}{a}e^{a\xi}\right) 
=\frac{\pi^2 a}{2\omega\sinh(\pi\omega/a)}
\delta(\omega-\omega') 
\label{I2}
\end{equation} 
for $ \omega, \omega' \in {\bf R_+}$. With this procedure we
reduce the emission amplitude (\ref{MAEM}) to 
\begin{eqnarray}
^M{\cal A}_{\mbox{\scriptsize em}} (k,{\bf k}) =  
\frac{1}{({4 \pi a k})^{1/2}}
& &\int_0^{+\infty }  
\frac{d\omega }{\sinh^{1/2} (\pi \omega/a)}\;
\left\{ 
\tilde \jmath_{R} (\omega , \omega , {\bf k}_\bot )
\left(\frac{k + k_x}{k - k_x} \right)^{- i\omega /2a}
e^{\pi \omega/2a}
\right.
\nonumber \\
& &
\left.
+ \tilde \jmath_{R} (-\omega , \omega , {\bf k}_\bot )
\left(\frac{k + k_x}{k - k_x}  \right)^{+i\omega /2a}
e^{- \pi \omega/2a} 
\right\} .
\label{MAEM3}
\end{eqnarray} 

In the left-hand side, we have the particle emission amplitude 
of Minkowski particles as computed by inertial observers. Hence, 
our next task will be to interpret $\tilde \jmath_R (\pm
\omega, \omega, {\bf k} _\bot)$ in terms of the emission
and absorption amplitudes of Rindler particles
\begin{equation} 
^R{\cal A}_{\mbox{\scriptsize em}}
= _R\langle  {\omega {\bf k}_\bot} \vert \; \hat{\cal S} \vert 0
\rangle_R ,\;\; ^R{\cal A}_{\mbox{\scriptsize abs}} 
= _R\langle 0 \vert \hat{\cal S} \vert \omega  {\bf k}_\bot \rangle_R
\label{acima}
\end{equation}
respectively, where $\hat{\cal S}$ is given in (\ref{S1}).
Using explicitly (\ref{S1}) and (\ref{FIXMU2}) in (\ref{acima})
we obtain 
\begin{equation}
\tilde \jmath_R (\omega, \omega, {\bf k}_\bot) = 
^R{\cal A}_{\mbox{\scriptsize em}} (\omega, {\bf k}_\bot) ,\;\;\;
\tilde \jmath_R (-\omega, \omega, {\bf k}_\bot) = 
^R{\cal A}_{\mbox{\scriptsize abs}} (\omega, - {\bf k}_\bot) .
\end{equation}
As a consequence, we can express the Minkowski particle emission
amplitude (\ref{MAEM3}) as 
\begin{eqnarray}
^M{\cal A}_{\mbox{\scriptsize em}} (k,{\bf k}) =  
\frac{1}{({4 \pi a k})^{1/2}}
&  & \int_0^{+\infty }  
\frac{d\omega }{\sinh^{1/2} (\pi \omega/a)}\;
\left\{ 
^R{\cal A}_{\mbox{\scriptsize em}} (\omega, {\bf k}_\bot)
\left(\frac{k + k_x}{k - k_x} \right)^{- i\omega /2a}
e^{\pi \omega/2a}
\right.
\nonumber \\
& &
\left.
+ ^R{\cal A}_{\mbox{\scriptsize abs}} (\omega, -{\bf k}_\bot)
\left(\frac{k + k_x}{k - k_x}  \right)^{+i\omega /2a}
e^{- \pi \omega/2a} 
\right\} .
\label{MAEM4}
\end{eqnarray}

Finally, we are ready to calculate the emission probability of Minkowski
particles and relate the result with the emission and absorption probability
of Rindler particles. For this purpose,
it is useful to introduce the following representation of
the delta function  
\begin{equation}
\frac{1}{2\pi a} \int_{-\infty}^{+\infty} \frac{dk_x}{k}
\left[\frac{k+k_x}{k-k_x}\right]^{-i(\omega -\omega')/2a}
= \delta (\omega -\omega') ,
\label{DR}
\end{equation} 
where $k=\sqrt{k_x^2 + k_\bot^2}$.
(This delta function representation can be cast in the more familiar
form
$ \int_{-\infty}^{+\infty} 
dK e^{-iK (\omega -\omega')}
= 2\pi \delta (\omega -\omega') 
$
after the change of variables $ k_x \to K =
\ln [(k+k_x)/(k-k_x)]$.)
Thus, introducing
(\ref{MAEM4}) in (\ref{PM}), and using (\ref{DR}) to perform
the integral in $k_x$ we are able to express the total emission rate of
Minkowski particles in terms of emission and absorption rates of 
Rindler particles in the following form   
\begin{equation}
^M{\cal P}_{\mbox{\scriptsize em}} =
^R{\cal P}_{\mbox{\scriptsize em}} + 
^R{\cal P}_{\mbox{\scriptsize abs}} ,
\label{PEM6}
\end{equation} 
where
\begin{equation}
^R{\cal P}_{\mbox{\scriptsize em}} =
\int d^{n-2} {\bf k}_\bot  \int_{0}^{+\infty} d\omega 
\vert  
^R{\cal A}_{\mbox{\scriptsize em}} (\omega, {\bf k}_\bot) 
\vert^2
\left( 1 + \frac {1}{e^{2\pi \omega /a} - 1} \right),
\label{FIM2}
\end{equation}
and
\begin{equation}
^R{\cal P}_{\mbox{\scriptsize abs}} =
\int d^{n-2} {\bf k}_\bot  \int_{0}^{+\infty} d\omega 
\vert ^R{\cal A}_{\mbox{\scriptsize abs}} (\omega, - {\bf k}_\bot)
\vert^2
\frac {1}{e^{2\pi \omega /a} - 1} .
\label{FIM3}
\end{equation}
The thermal factors which appear in (\ref{FIM2}) and (\ref{FIM3})
are in agreement with the fact that the Minkowski vacuum
corresponds to a thermal state with respect to uniformly
accelerated observers \cite{U}. The physical content of 
(\ref{PEM6}) combined with (\ref{FIM2}) and (\ref{FIM3})  
can be summarized as follows: {\em The emission of a
Minkowski particle in the vacuum with some fixed transverse 
momentum ${\bf k}_\bot$
as described by an inertial observer {\em (\ref{PM})} will correspond 
either to the emission of a Rindler particle with 
the same transverse momentum ${\bf k}_\bot$, or to the
absorption of a Rindler particle with transverse momentum 
$-{\bf k}_\bot$ from the Davies-Unruh thermal bath
as described by uniformly accelerated observers.} 
This is consistent with energy conservation arguments, since 
a source following an arbitrary trajectory is not static neither in the
inertial nor in the accelerated frame.
Notice that the 
conservation of transverse momentum in both frames appears 
naturally enclosed in the result above. 
This is mandatory since the
transverse momentum is invariant under boosts. 
Recently, Eq. (\ref{PEM6}) was used in the particular
case of uniformly accelerated Unruh--DeWitt detectors 
kept switched on for a finite period of time \cite{HMP}. 
Indeed, the excitation of finite--time Unruh-DeWitt detectors 
in the inertial vacuum is associated 
with the emission of a Minkowski particle as described by inertial
observers and with the absorption or emission of a Rindler particle
as described by uniformly accelerated observers. 
We have shown here that this is
also true for a general class of detectors 
following arbitrary worldlines.
In the particular case where the
source is a uniformly accelerated Unruh-DeWitt detector, we obtain that
$^R{\cal A}_{\mbox{\scriptsize em}}=0$, which implies
that the absorption of a Rindler particle corresponds
uniquely to the emission of a Minkowski particle \cite{UW}.
Although this particular case is presently being revisited, we
argue that the interpretation given above for Eq. (\ref{PEM6}) 
must be free of dispute for 
sources following realistic trajectories in the Rindler wedge.  
It is interesting that were we uniformly accelerated observers, 
it would be  (\ref{FIM2}) and (\ref{FIM3}), rather than (\ref{PM}) 
the proper expressions to be used in order to describe our physical 
experience.

\begin{flushleft}
{\bf{\large Acknowledgements}}
\end{flushleft}

I am really indebted to Atsushi Higuchi for various
enlightening discussions. I am also very grateful to 
Ulrich Gerlach for carefully reading a previous version of this
manuscript.   
This work was partially supported by Conselho Nacional de
Desenvolvimento Cient\'\i fico e Tecnol\'ogico.


\newpage

\begin{thebibliography}{99}

\bibitem{Gr} P.G.\ Grove,   Class.\ Quant.\ Grav.\ {\bf 3} (1986) 801.

\bibitem{RSG} D.J.\ Raine, D.W.\ Sciama, and  P.G.\ Grove,  Proc.\ R.\ Soc.\ 
London {\bf 435} (1991) 205.

\bibitem{U2} W.G.\ Unruh, Phys.\ Rev.\ D\ {\bf 46} (1992) 3271.

\bibitem{MPB} S.\ Massar, R.\ Parentani, and R.\ Brout, 
Class.\ Quant.\ Grav.\ {\bf 10} (1993) 385.

\bibitem{H} F.\ Hinterleitner, Ann.\ Phys.\ (N.Y.) {\bf 226} (1993) 165.

\bibitem{AM} J.\ Audretsch and R.\ M\"uller, Phys.\ Rev.\ D {\bf 49} 
(1994) 6566.

\bibitem{UW} W.G.\ Unruh and R.M.\ Wald, Phys.\ Rev.\ D {\bf 29} (1984) 1047. 

\bibitem{U} W.G.\ Unruh,   Phys.\ Rev.\ D {\bf 14} (1976) 870.

\bibitem{D} B.S.\ DeWitt, {\it General Relativity}, eds. S.W.\ Hawking 
 and W.\ Israel   (Cambridge University Press, Cambridge, 1979).

\bibitem{FR} T.\ Fulton and F.\ Rohrlich, Ann.\ Phys.\ (N.Y.) {\bf 9} (1960)
499.

\bibitem{HMP} A.\ Higuchi, G.E.A.\ Matsas  and C.B.\ Peres,   
Phys. Rev. D {\bf 48} (1993) 3731. 

\bibitem{AMH}  J.\ Audretsch, R.\ M\"uller and M.\ Holzmann,
Class.\ Quant.\ Grav.\ {\bf 12} (1995) 2927.
 
\bibitem{BD} N.D.\ Birrell and P.C.W.\ Davies, {\it Quantum Field
Theory in Curved Space}, (Cambridge University Press, Cambridge, 1982).

\bibitem{M} G.E.A.\ Matsas, Gen.\ Relat.\ Grav.\ {\bf 26} (1994) 1165.

\bibitem{F} S.A.\ Fulling,  Phys. Rev. D {\bf 7} (1973) 2850. 

\bibitem{HM} A.\ Higuchi and G.E.A.\ Matsas  Phys. Rev. D {\bf 48} (1993)
689.  

\bibitem{GR} I.S.\ Gradshteyn and I.M.\ Ryzhik, {\it Table of Integrals,
Series and Products} (Academic Press, New York, 1980).

\bibitem{G} U.\ Gerlach,  Phys. Rev. D {\bf 38} (1988) 514. 
 
\end{thebibliography}
\end{document}